\documentclass{emulateapj}
\input{colordvi.tex}
\usepackage{natbib}
\usepackage{apjfonts}
\usepackage{amsmath}
\usepackage{graphicx}

\shorttitle{Detectability of Pair Echos from GRBs}
\shortauthors{Takahashi et al.}

\begin{document}

\title{Detectability of Pair Echos from Gamma-Ray Bursts and Intergalactic Magnetic Fields}

\author{Keitaro Takahashi\altaffilmark{1,*},
        Kohta Murase\altaffilmark{1},
        Kiyotomo Ichiki\altaffilmark{2},
        Susumu Inoue\altaffilmark{3},
        Shigehiro Nagataki\altaffilmark{1}
}

\altaffiltext{1}{Yukawa Institute for Theoretical Physics,
Kyoto University, Oiwake-cho, Kitashirakawa, Sakyo-ku, Kyoto,
606-8502, Japan}
\altaffiltext{2}{Research Center for the Early Universe,
University of Tokyo, 7-3-1 Hongo, Bunkyo-ku, Tokyo, 113-0033, Japan}
\altaffiltext{3}{Department of Physics,
Kyoto University, Oiwake-cho, Kitashirakawa, Sakyo-ku, Kyoto,
606-8502, Japan}
\altaffiltext{*}{keitaro@yukawa.kyoto-u.ac.jp}

\begin{abstract}
High-energy emission from gamma-ray bursts (GRBs) can give rise to
pair echos, i.e. delayed inverse Compton emission from secondary
$e^{\pm}$ pairs produced in $\gamma-\gamma$ interactions with
intergalactic background radiation. We investigate the detectability
of such emission with modern-day gamma-ray telescopes.
The spectra and light curves are calculated for a wide range of parameters,
applying the formalism recently developed by Ichiki et al.
The flux depends strongly on the unknown magnitude and coherence
length of intergalactic magnetic fields, and we delineate the range
of field strength and redshift that allow detectable echos.
Relevant uncertainties such as the high-energy cutoff of
the primary gamma-ray spectrum and the intensity of the cosmic
infrared background are addressed. GLAST and MAGIC may be able
to detect pair echo emission from GRBs with redshift $\lesssim 1$
if the primary spectra extend to $\sim 10 ~ {\rm TeV}$.
\end{abstract}

\keywords{magnetic fields ---  gamma rays: bursts ---
radiation mechanisms: nonthermal}

\section{Introduction}

Gamma-ray bursts (GRBs) are expected to be emitters of high-energy
gamma-rays, possibly up to TeV energies and above. Such TeV photons
can interact with photons of the cosmic infrared background to
produce electron-positron pairs, which in turn generate secondary
inverse Compton gamma-rays in the $(1 - 100) ~ {\rm GeV}$ range
that arrive with a characteristic time delay. Hereafter we shall
call such delayed secondary emission "pair echos". Pair echo emission
has attracted much attention as a powerful tool to probe
high-energy emission of GRBs because its detection would directly
prove that the spectrum of the GRB emission extends to TeV range.

So far, the observations of high-energy gamma-rays from GRBs have
been limited. There are some detections of GeV gamma-rays by EGRET 
\citep{1994ApJ...422L..63S,1994Natur.372..652H},
while only upper bounds have been obtained for TeV range
\citep{2006ApJ...641L...9A,2007ApJ...666..361A},
although some tentative detections have been claimed
\citep{1996A&A...311..919T,1998A&A...337...43P,2000ApJ...533L.119A}.
Thus, it is not clearly known observationally whether the spectrum
extends to TeV region. On the other hand, high energy emission of
GRBs has been studied theoretically by many authors and the spectrum
depends highly on the emission mechanism and physical parameters. 
In the conventional internal shock model where prompt emission is
attributed to synchrotron radiation from electrons, TeV gamma rays
can be produced via leptonic mechanisms 
\citep{1996ApJ...471.L91P,2003ApJ...585..885G,2007ApJ...671..645A} 
and/or hadron-related mechanisms 
\citep{1998ApJ...509L..81T,2004ApJ...613..448P,2004A&A...418L...5D,
2007ApJ...671..645A,2008arXiv0801.2861M}.
Although these high-energy gamma rays can suffer from the pair-creation
process within the source, they can escape from the source if the
emission radius and/or bulk Lorentz factor are large enough
\citep{2004ApJ...613.1072R,2008ApJ...676.1123M}. Hence, as long as
the conventional internal shock model is valid, we may expect that
at least some fraction of GRBs have spectra extended to TeV region. 

\cite{1995Natur.374..430P} proposed that pair echo emission can also
probe intergalactic magnetic fields because the delay time and
the flux depends strongly on their magnitude and coherence length.
Duration and spectra of pair echos from GRBs including the effects
of magnetic fields have been studied in subsequent works 
\citep{2002ApJ...580.1013D,2004ApJ...604..306W,
2004ApJ...613.1072R,2004MNRAS.354..414A,2007ApJ...656..306C,
2007ApJ...671.1886M} and revealed that intergalactic magnetic fields
as tiny as $\sim 10^{-20} ~ {\rm G}$ could be probed by its observation.
In our previous work \citep{2007arXiv0711.1589I}, we constructed
an analytic model which properly incorporates geometrical effects
of particle paths on time delay. It allow us to calculate the time
evolution of pair echos more precisely than other simple
analyses presented by \citet{2002ApJ...580L...7D}.

Plaga's method is currently the only way to probe such tiny magnetic
fields, while other methods utilizing Faraday rotation or cosmic
microwave background are sensitive to magnetic fields of order
$1 ~ {\rm nG}$. Actually, presence of tiny intergalactic magnetic
fields have been predicted by several mechanism, such as inflation
\citep{1988PhRvD..37.2743T,2007JCAP...02...30B},
reionization
\citep{2000ApJ...539..505G,2005A&A...443..367L} and density fluctuations
\citep{2005PhRvD..71d3502M,2005PhRvL..95l1301T,2006Sci...311..827I,
2007arXiv0710.4620T,2008arXiv0805.0169M},
and observation of intergalactic magnetic fields would give important
information on the origin of galactic magnetic fields
\citep{2002RvMP...74..775W}.

It would be important to argue the detectability of pair echo
emission from GRBs for ongoing ground-based instruments such as
MAGIC, which we take as a representative of this type, H.E.S.S.,
VERITAS, CANGAROO III and MILAGRO, and an upcoming satellite GLAST.
There are several factors concerning the detectability:(1) the object
must emit TeV-gamma-rays which lead to pair echo emission, (2) the flux
of pair echos must be higher than detector sensitivity,
(3) the pair echo emission must not be masked by other emission such as
afterglow. Concerning the point (1), we just assume the cutoff energy,
$E_{\rm cut}$, of the primary spectrum. Our main subject is
the point (2). The flux of pair echo emission depends on, other
than the distance to the GRBs and properties of magnetic fields,
the amount of absorption of primary gamma-rays. It is determined
by the spectrum of primary emission and the amount of CIB.
We estimate, using our formalism to calculate spectra and light curves
of pair echos, the detectability in
(redshift)-(magnitude of magnetic fields) plane varying $E_{\rm cut}$
and CIB model.

Concerning the point (3), it should be noted that afterglows can
also have high energy emission above GeV \citep{2001ApJ...559..110Z,
2001ApJ...546L..33W,2008MNRAS.384.1483F}, although it has not been
confirmed by observation either. Typically afterglow continues
for several days so that its high energy emission can mask
the pair echos. There have been many theoretical predictions
on high-energy afterglow, but we postpone the comparison of spectra
and lightcurves between afterglow and pair echos and give
some comments in the last section.

In section 2, we summarize the basic elements of pair echo emission
and the calculation formalism constructed in \cite{2007arXiv0711.1589I},
and apply the formalism to GRB cases to show the spectra and light curves
for a wide range of parameters. Then, we argue the detectability of
pair echo emission considering uncertainties in high-energy cutoff
in primary spectrum and the density of cosmic infrared background
in section 3. Finally we give a summary in section 4.

\section{pair echo emission \label{section:delayed-emission}}

Let us summarize the basic elements of pair echo emission from
high-energy gamma-ray sources. A gamma-ray with energy
$E_{\gamma} \gtrsim 1 ~ {\rm TeV}$ emitted from a source,
called a primary gamma-ray, pair-annihilates with an ambient infrared
photon to create a electron-positron pair with the mean free path,
$
\lambda_{\gamma \gamma}
= 1/(0.26 \sigma_T n_{\rm IR})
= 19 ~ {\rm Mpc} ( n_{\rm IR}/0.1 ~ {\rm cm}^{-3} )^{-1},
$
where $\sigma_T$ is the Thomson cross section and $n_{\rm IR}$ is
the number density of infrared background. Created charged particles
(electron or positron) have energies $E_e = E_{\gamma}/2$ and
upscatter ambient CMB photons
to produce high-energy secondary gamma-rays. The average energy of
upscattered photons is,
$
\langle E_{\rm delay} \rangle
= 2.7 T_{\rm CMB} \gamma_e^2
= 2.5 ~ {\rm GeV}
  ( E_{\gamma}/2 ~ {\rm TeV} )^2,
$
where $\gamma_e = E_e/m_e$ is the Lorentz factor of a charged particle
and $T_{\rm CMB} = 3 ~ {\rm K}$ is the CMB temperature. We can see
that, considering the energy range of the primary gamma-rays
as $1 \sim 10 ~ {\rm TeV}$, the typical energy range of pair echo
emission is $1 \sim 100 ~ {\rm GeV}$.
The mean free path of charged particles is evaluated as,
$
\lambda_{\rm IC, scat}
= 1/(\sigma_T n_{\rm CMB}) = 1.2 ~ {\rm kpc},
$
where $n_{\rm CMB} \approx 420 ~ {\rm cm}^{-3}$ is the number density
of CMB. The charged particles scatter CMB photons successively
until they lose most of their energy after propagating the cooling length,
$
\lambda_{\rm IC, cool}
= 3 m_e^2/(4 E_e \sigma_T U_{\rm CMB})
= 350 ~ {\rm kpc} ( E_e/1 ~ {\rm TeV} )^{-1},
$
where $U_{\rm CMB}$ is the CMB energy density. Note that ambient infrared
photons should also be the target of IC scattering. They can be important
for pair echo gamma-rays with $E_{\rm delay} \gtrsim 10 ~ {\rm GeV}$
and/or even lower energies at late time \citep{2007ApJ...671.1886M}.

Pair echo emission is induced by angular spreading and magnetic fields.
The direction of upscattered photons deviates from the original directions
of the charged particle and primary gamma-ray by an angle $\sim 1/\gamma_e$
so that the typical delay time can be evaluated as,
\begin{eqnarray}
\Delta t_{\rm ang}
&\sim& \frac{1}{2 \gamma_e^2}
       (\lambda_{\gamma \gamma} + \lambda_{\rm IC, cool})
\nonumber \\
&\sim& 3 \times 10^2 ~ {\rm sec}
       \left( \frac{E_{\rm delay}}{1 ~ {\rm GeV}} \right)^{-1}
       \left( \frac{n_{\rm CIB}}{0.1 ~ {\rm cm}^{-3}} \right)^{-1}.
\end{eqnarray}
If magnetic fields are present in the region of the propagation of
charged particles, their directions are further deflected by
magnetic fields. For weak magnetic fields with coherence length
$r_{\rm coh} (< \lambda_{\rm IC, cool})$, the variance of the 
deflection angle due to magnetic fields is,
$
\sqrt{\langle \theta_{\rm B, random}^2 \rangle}
= \sqrt{\lambda_{\rm IC, cool}/6 r_{\rm coh}}
  r_{\rm coh}/r_{\rm L}
= 2.2 \times 10^{-6}
  ( E_e/1 ~ {\rm TeV} )^{-3/2}
  ( B/10^{-18} ~ {\rm G} )
  ( r_{\rm coh}/100 ~ {\rm pc} )^{1/2},
$
where $r_{\rm L}$ is the Larmor radius,
$
r_{\rm L}
= 1.1 ~ {\rm Gpc} ( E_e / 1 ~ {\rm TeV} )
  ( B / 10^{-18} ~ {\rm G} )^{-1}.
$
Then typical delay time due to magnetic deflection is,
\begin{eqnarray}
\Delta t_{\rm B}
&=& \frac{1}{2} (\lambda_{\gamma \gamma} + \lambda_{\rm IC, cool})
    \langle \theta_{\rm B, random}^2 \rangle
\nonumber \\
&\approx&
  2 \times 10^4 ~ {\rm sec}
  \left( \frac{E_{\rm echo}}{1 ~ {\rm GeV}} \right)^{-3/2}
  \left( \frac{B}{10^{-18} ~ {\rm G}} \right)^2
\nonumber \\
& & \times
  \left( \frac{r_{\rm coh}}{100 ~ {\rm pc}} \right)
  \left( \frac{n_{\rm CIB}}{0.1 ~ {\rm cm}^{-3}} \right)^{-1}.
\label{eq:delay-time-B-random}
\end{eqnarray}
Eventually, the true delay time is
$\Delta t = {\rm max}[\Delta t_{\rm ang}, \Delta t_{\rm B}]$
and we can probe magnetic field by measuring
the time delay between primary and secondary emissions.
Because the mean free path of the primary gamma-rays can be larger
than the typical sizes of galaxy and cluster of galaxies, the 
primary gamma-rays will escape from the host galaxy
/cluster of galaxies of the GRB before they pair-annihilate with ambient
infrared photons. Therefore, the propagation region of
charged particles will mostly be in void region and this
is where we can expect to probe magnetic fields by our method.

In our previous study \citep{2007arXiv0711.1589I}, the spectrum
and light curves of pair echo from a GRB was investigated and
let us summarize the basic formalism here. For a primary gamma-ray
fluence $dN_{\gamma}/dE_{\gamma}$, the time-integrated flux of 
charged particles over the GRB duration
can be written as,
\begin{equation}
\frac{dN_{e,{\rm 0}}}{d\gamma_e} (\gamma_e)
= 4 m_e
  \frac{dN_{\gamma}}{dE_{\rm \gamma}}(E_{\gamma} = 2 m_e \gamma_e)
  \left[1-e^{-\tau_{\gamma \gamma}(E_{\gamma} = 2 \gamma_e m_e)}\right].
\label{eq:dN0dgamma}
\end{equation}
Note that we shall assume the GRB duration $T^{\prime}=50$ s in this letter.
Here $\tau_{\gamma \gamma}(E_\gamma)$ is the optical depth to pair production
for gamma-rays with energy $E_\gamma$. From this,
the spectrum of the pair echo emission can be calculated as
\begin{equation}
\frac{d^2 N_{\rm echo}}{dt dE_\gamma}
= \int d\gamma_e \frac{dN_e}{d{\gamma_e}}
  \frac{d^2 N_{\rm IC}}{dt dE_\gamma}.
\end{equation}
where $d^2 N_{\rm IC}/dt dE_\gamma$ is the IC power from
a single electron. Here $dN_e/d{\gamma_e}$ is the total
time-integrated flux of charged particles responsible for
the pair echo emission observed at time $t_{\rm obs}$ after
the burst and is nontrivially related to $dN_{e, \rm 0}/d{\gamma_e}$
in Eq. (\ref{eq:dN0dgamma}). The relation was calculated in
\cite{2007arXiv0711.1589I} taking geometrical effects due to
random walks properly into account, which makes it possible
to follow the time evolution of pair echo emission. Calculation
needs numerical integration but it can be written roughly as,
$dN_e/d{\gamma_e} =
 (\lambda_{\rm IC, cool}/c \Delta t) dN_{e,{\rm 0}}/d\gamma_e$,
as given by \cite{2002ApJ...580L...7D}.

\begin{figure}[t]
\epsscale{1.1}
\plotone{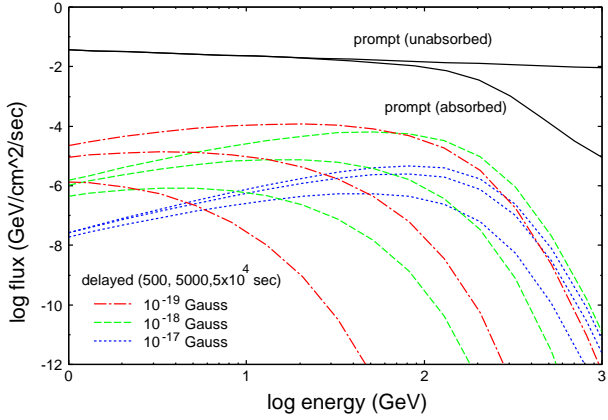}
\caption{Typical spectra of pair echo emission at
$t_{\rm obs} = 500, 5000$ and $5 \times 10^4 ~ {\rm sec}$ from
a GRB located at $z=0.3$. The coherence length of magnetic fields
are set to $r_{\rm coh} = 100 ~ {\rm pc}$. Other parameters
are shown in the text.}
\label{fig:spectrum}
\end{figure}

\begin{figure}[t]
\epsscale{1.1}
\plotone{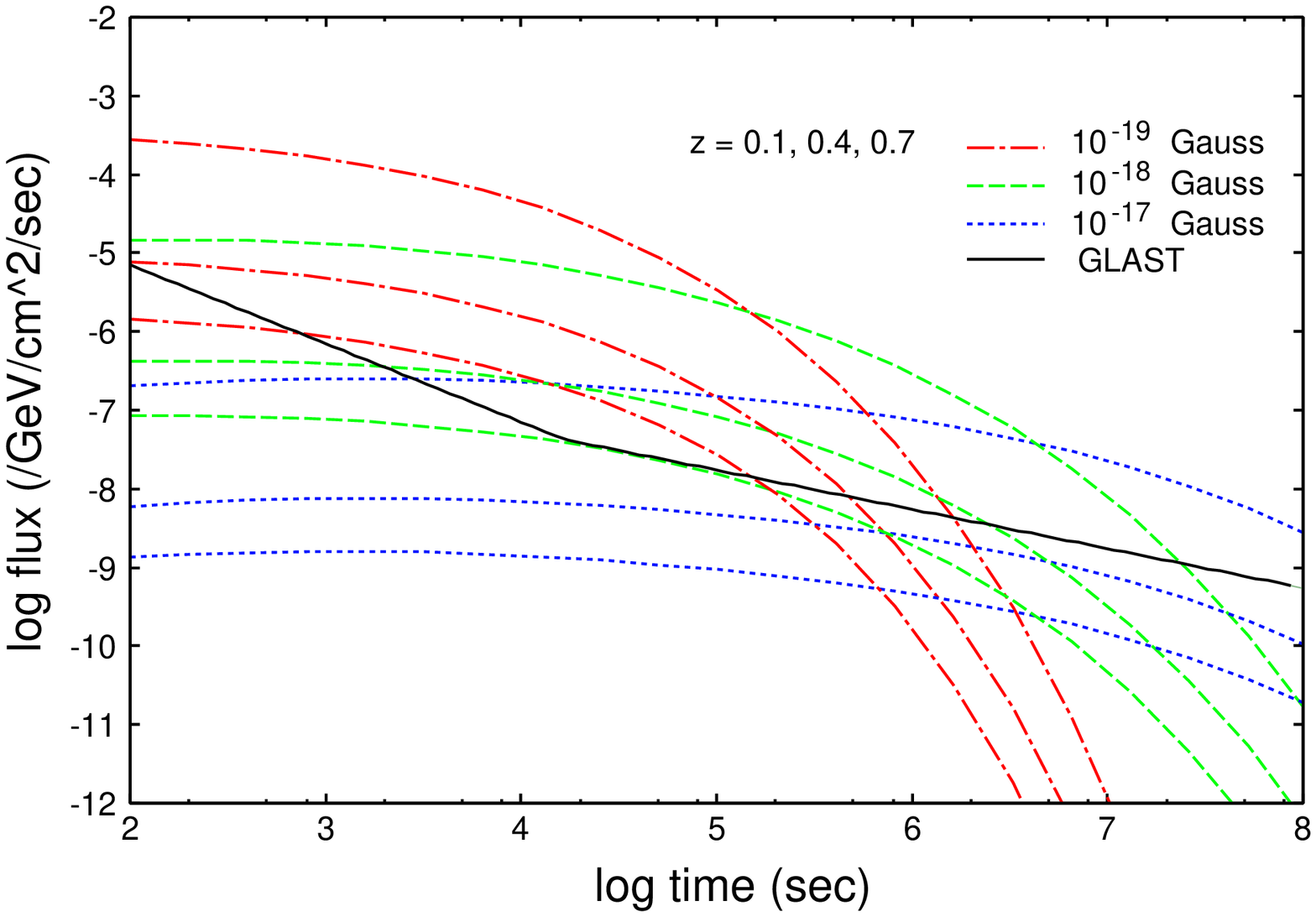}
\plotone{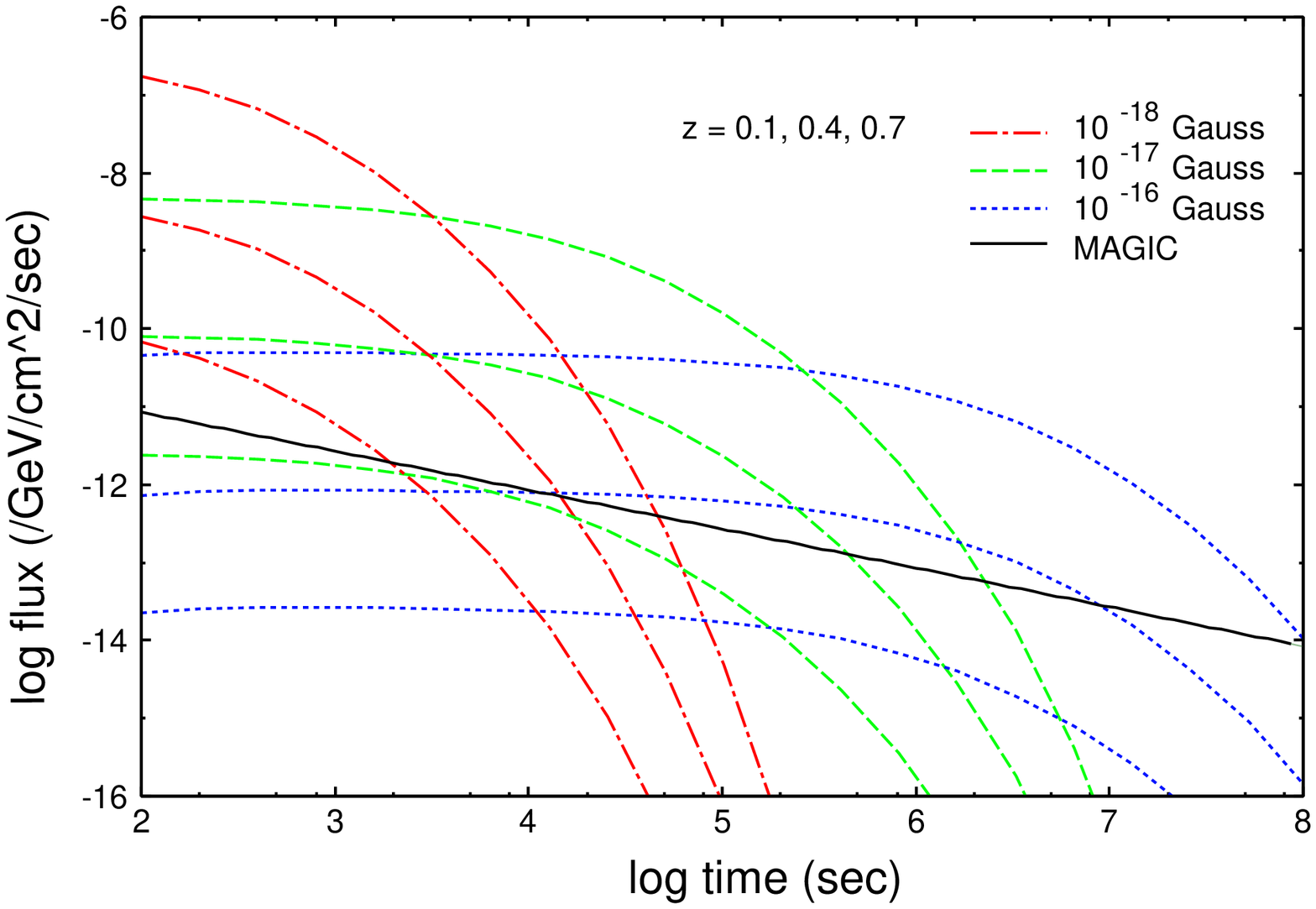}
\caption{Typical light curves of pair echo emission at
$E_{\rm delay} = 1 ~ {\rm GeV}$ (above)
and $E_{\rm delay} = 100 ~ {\rm GeV}$ (below)
from GRBs located at $z=0.1, 0.4$ and $0.7$. The coherence 
length of magnetic fields are fixed to
$r_{\rm coh} = 100 ~ {\rm pc}$.
Sensitivities of GLAST at $1 ~ {\rm GeV}$ (above) and MAGIC
at $100 ~ {\rm GeV}$ (below) are also shown.}
\label{fig:lightcurve}
\end{figure}

Typical spectra and light curves of pair echo emission from GRBs
are shown in Figs. \ref{fig:spectrum} and \ref{fig:lightcurve}.
Here we assumed that the primary gamma-ray flux is power-law with
index $2.2$, that is, $dN_{\gamma}/dE_{\gamma} \propto E_{\gamma}^{-2.2}$,
for $0.1 ~ {\rm TeV} < E_{\gamma} < E_{\rm cut} = 10 ~ {\rm TeV}$
with isotropic energy 
$E_{\gamma,[0.1,10]}^{\rm{iso}} = 3 \times 10^{53} ~ {\rm erg}$
similarly to \cite{2007arXiv0711.1589I}.
Such a large value may be rather optimistic \citep{2006ApJ...641L...9A}, but 
possible if the strong synchrotron self-inverse Compton emission 
occurs. In this letter we shall adopt this value for demonstrations,
and see, e.g., \cite{2007ApJ...671.1886M} for more conservative cases. 
As to the density of ambient infrared photons, we adopted
the "best fit" model developed by 
\cite{2002A&A...386....1K,2004A&A...413..807K} as a fiducial model.
The sensitivities of GLAST and MAGIC are also shown in
Fig. \ref{fig:lightcurve}, respectively.
As can be seen, the flux of pair echos decays exponentially
with time scale estimated in Eq. (\ref{eq:delay-time-B-random}).

It is important to note that the total fluence of pair echo
emission is determined by the amount of absorbed primary gamma-rays
and does not depend on the properties of magnetic field.
On the other hand, the flux of pair echo emission is roughly
the fluence divided by $\Delta t$ and does depend on
magnetic fields. From this fact, we see that there is
an upper bound for the amplitude of magnetic field probed
by our method. This is because strong magnetic fields lead
to small flux of pair echo emission and make observations
more difficult. The maximum measurable amplitude depends on several
factors such as the distance of the source and detector
sensitivity, and turns out to be typically about
$10^{-16} ~ {\rm G}$ as we see later. Contrastingly,
if magnetic fields are too weak, the time delay is dominated
by angular spreading or hidden by the GRB duration, when we cannot 
obtain information on magnetic fields from pair echo emissions.
The minimum measurable amplitude can be read from the ratios
of delay times due to angular spreading and magnetic fields,
\begin{equation}
\frac{\Delta t_{\rm A}}{\Delta t_{\rm B}}
= 1.5 \times 10^{-2}
    \left( \frac{E_{\rm echo}}{1 ~ {\rm GeV}} \right)^{1/2}
    \left( \frac{B}{10^{-18} ~ {\rm G}} \right)^{-2}
    \left( \frac{r_{\rm coh}}{100 ~ {\rm pc}} \right)^{-1}.
\end{equation}
Depending on the coherence length, the minimum measurable amplitude
is expected to be $10^{-19} \sim 10^{-20} ~ {\rm G}$.
However, it should be noted that pair echo emission itself is
easier to observe for weaker magnetic fields. Thus, we have
both an upper bound and a lower bound for the measurable amplitude.

\section{Detectability of pair echo emission from GRBs
\label{section:detectability}}

\begin{figure}[t]
\epsscale{1.1}
\plotone{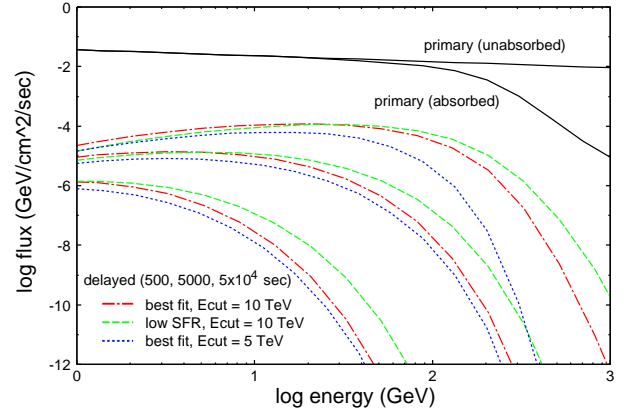}
\caption{Comparison spectra of pair echo emission from a GRB
located at $z = 0.3$ varying CIB model and $E_{\rm cut}$.
The magnitude and coherence length of magnetic fields are set
to $B = 10^{-18} ~ {\rm Gauss}$ and $r_{\rm coh} = 100 ~ {\rm pc}$,
respectively. Other parameters are shown in the text.}
\label{fig:spectrum_compare}
\end{figure}

As we stated in section 1, there are several factors which determine
the flux of pair echo emission. First of all, as we saw in the previous
section, strong magnetic fields and large coherence length lead to
large time delay and hence low flux. Next, the density of CIB determines
the amount of absorption of primary gamma-rays and then the flux of
pair echos, but we have neither enough observational data nor
theoretical understanding about it.
Kneiske et al. \citep{2002A&A...386....1K,2004A&A...413..807K}
constructed several semi-empirical models of CIB evolution which well
fit the observational data (see also \cite{2006ApJ...648..774S}).
In Fig. \ref{fig:spectrum_compare}, we show the spectra of pair echos
for their "low SFR" model, which predicts lower CIB density,
compared to those for "best fit" model. 
There are two competing
effects to understand the behavior of the spectra. Lower CIB density
results in less absorption of primary gamma-rays and less pair echo
emission, while re-absorption of pair echo emission, which is important
at high energies, is also reduced. In fact, contributions from
upscattered CIB photons can also bring further complication
\citep{2007ApJ...671.1886M}. 

Yet another uncertainty is $E_{\rm cut}$. We show the spectra
of pair echo emission for $E_{\rm cut} = 5 ~ {\rm TeV}$ also
in Fig. \ref{fig:spectrum_compare}. As one can understand from
the typical relation between the energies of primary and pair
echo gamma-rays,
$
\langle E_{\rm delay} \rangle
= 2.7 T_{\rm CMB} \gamma_e^2
= 2.5 ~ {\rm GeV}
  ( E_{\gamma}/2 ~ {\rm TeV} )^2,
$
in the absence of high-energy primary gamma-rays, high-energy
pair echo emission is substantially reduced at early phase.
Contrastingly, dependence on $E_{\rm cut}$ is rather weak
at late phase because late-phase pair echo emission is induced
by low-energy primary gamma-rays.

\begin{figure}[t]
\epsscale{1.1}
\plotone{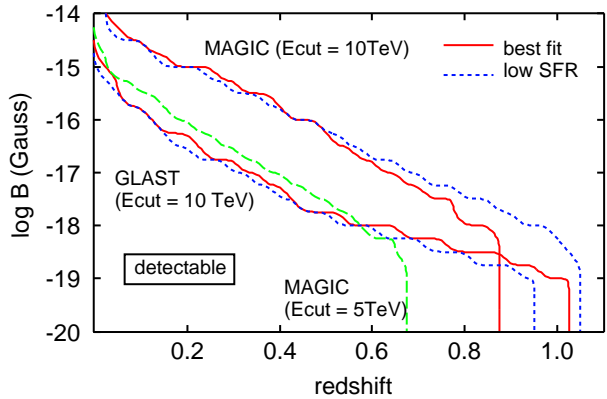}
\caption{Detectable region of pair echo emission in $z-B$ plane
with GLAST and MAGIC for tangled magnetic fields with
$r_{\rm coh} = 100 ~ {\rm pc}$. Pair echo emission is considered
to be detectable if the flux exceeds the detector sensitivity
during the time between $10^2 \sim 10^7 ~ {\rm sec}$ after the burst.
Solid and dotted lines are estimations with "best fit" and "low SFR"
models of CIB, respectively. Dashed line is for MAGIC with
$E_{\rm cut} = 5 ~ {\rm TeV}$ and "best fit" CIB model.
}
\label{fig:B-z_random}
\end{figure}

In Fig. \ref{fig:B-z_random} we show regions in $z-B$ plane where
the flux of pair echo emission exceeds the detector sensivities during
the time interval $10^2 ~ {\rm sec} \sim 10^7 ~ {\rm sec}$, varying
CIB model and $E_{\rm cut}$. From Eqs. (\ref{eq:delay-time-B-random})
we see that the delay time is shorter for higher energies so that MAGIC
can probe stronger magnetic fields than GLAST. Below
$B \sim 10^{-19} ~ {\rm G}$, the detectability does not depend on
$B$ because the delay time is dominated by angular spreading or hidden
by the GRB duration, and the flux above $\sim$ GeV energies does not 
depend on $B$. Dependence on $E_{\rm cut}$
is strong for MAGIC because its energy range is relatively high
($\gtrsim 100 ~ {\rm GeV}$). In contrast, GLAST detects much lower-energy
gamma-rays ($\lesssim 10 ~ {\rm GeV}$) and is little affected by
$E_{\rm cut}$ as long as it is high enough. Therefore, the detectability 
curve is not shown for $E_{\rm cut} = 5 ~ {\rm TeV}$, but of cource it
should be affected for even lower $E_{\rm cut}$
\citep{2007ApJ...671.1886M}. Thus, we can see in the figure
how large and small fields can be probed with GRBs with various redshift.
For our specific GRB parameters which might be optimistic, the
redshift must be rather small, $z \lesssim 1.0$, to detect pair echo
emission. Since about $10 \%$ of GRBs will be located
with $z \lesssim 1.0$, we expect that there would be some 
possibilities to probe magnetic fields with GRBs.

\section{Discussion and Summary \label{section:summary}}

In this paper, we investigated the detectability of pair echo emission
from gamma-ray bursts for GLAST and MAGIC. We calculated the spectra
and light curves of pair echo emission using the formalism developed
in \citep{2007arXiv0711.1589I} varying intergalactic magnetic fields,
CIB model and cutoff energy. We showed the detectable region in
(redshift)-(magnitude) plane and found that GLAST and MAGIC
would be able to detect pair echo emission from GRBs with redshift
$\lesssim 1$, while the cutoff energy will affect the detectability
substantially for MAGIC. Thus, relatively nearby GRBs would allow us
to probe high-energy emission of GRBs and properties of intergalactic
magnetic fields.

One important ingredient which we did not discuss in detail is
potential masking of pair echo emission by high-energy component
of afterglow. While afterglows decays like power-law with time,
pair echo emission lasts for a characteristic time determined by
magnetic fields and then decays exponentially. Therefore, pair echo
emission would tend to dominate afterglow at late phase, if
intergalactic magnetic fields are relatively strong.
As was pointed out in section 2, high-energy component of pair echo
emission at late phase has a contribution from upscattered CIB.
Thus, if pair echo emission at early phase would be masked by
afterglow, the CIB contribution would become important in
the argument of the detectability. This subject will be discussed
in our future work.

\acknowledgments

The works of KT, KM and KI are supported by a Grant-in-Aid for
the JSPS fellowship. The work of SN is supported by Grants-in-Aid
for Scientific Research of the Japanese Ministry of Education,
Culture, Sports, Science, and Technology 19104006, 19740139, 19047004.
The numerical calculations were carried out on Altix3700 BX2 at YITP
in Kyoto University.

\bibliographystyle{hapj}
\bibliography{grb}

\end{document}